\documentclass[pdflatex,sn-standardnature]{sn-jnl}

\usepackage{epstopdf}
\graphicspath{{./}}
\DeclareGraphicsExtensions{.pdf,.jpeg,.png,.eps,.gif}

\usepackage{times}
\usepackage{float}
\usepackage{caption}

\usepackage[caption=false,font=footnotesize]{subfig}

\usepackage{ieeetrantools}
\usepackage{mathtools}
\usepackage{array}

\theoremstyle{plain}
\newtheorem{thm}{Theorem}

\theoremstyle{definition}

\newtheorem{axiom}{Axiom}

\usepackage{xfrac}

\definecolor{shadecolor}{gray}{0.9}

\usepackage{glossaries}%

\setacronymstyle{long-sc-short}
\newacronym{pid}{PID}{proportional-integral-derivative controller}
\newacronym{bmf}{BMF}{Standard Binomial Filter(s)}
\newacronym{udbmf}{UDBMF}{Uniformly-Damped Binomial Filter}
\newacronym{sff}{SFF}{Somefun Filter}
\newacronym{udbmf5}{UDBMF5}{Five-percent Uniformly-Damped Binomial Filter}
\newacronym{bmf0}{UDBMF0}{Zero-percent Uniformly-Damped Binomial Filter}
\newacronym{bwf}{BWF}{Butterworth Filter}
\newacronym{sgf}{SGF}{Savitzky-Golay Filter}
\newacronym{cmac}{CMAC}{Critical Monotonic Amplitude Characteristic }

\usepackage{cleveref}

\jyear{2021}%

\theoremstyle{thmstyleone}%
\theoremstyle{thmstyletwo}%

\theoremstyle{thmstylethree}%
\newtheorem{definition}{Definition}%

\begin{document}

\title[Uniformly-Damped Binomial Filters]{Uniformly-Damped Binomial Filters: Five-percent Maximum Overshoot Optimal Response Design}


\author*[1]{\fnm{Oluwasegun} \sur{Somefun}}\email{oasomefun@futa.edu.ng}

\author[2]{\fnm{Kayode} \sur{Akingbade}}\email{kfakingbade@futa.edu.ng}
\equalcont{These authors contributed equally to this work.}

\author[1]{\fnm{Folasade} \sur{Dahunsi}}\email{fmdahunsi@futa.edu.ng}
\equalcont{These authors contributed equally to this work.}

\affil*[1]{\orgdiv{Department of Computer Engineering}, \orgname{ Federal University of Technology Akure}, \orgaddress{\postcode{PMB 704}, \state{Ondo}, \country{Nigeria}}}

\affil[2]{\orgdiv{Department of Electrical and Electronics Engineering}, \orgname{ Federal University of Technology Akure}, \orgaddress{\postcode{PMB 704}, \state{Ondo}, \country{Nigeria}}}



\abstract{In this paper, the five-percent maximum overshoot design of uniformly-damped binomial filters (transfer-functions) is introduced. First, the butterworth filter response is represented as a damped-binomial filter response. To extend the maximum-overshoot response of the second-order butterworth to higher orders, the binomial theorem is extended to the uniformly-damped binomial theorem. It is shown that the five-percent uniformly-damped binomial filter is a compromise between the butterworth filter and the standard binomial filter, with respect to the filter-approximation problem in the time and frequency domain. Finally, this paper concludes that in applications of interest, such as step-tracking, where both strong filtering and a fast, smooth transient-response with negligible overshoot are desired, the response of the normalized five-percent uniformly-damped binomial form is a candidate replacement for both the butterworth and standard binomial filter forms.}




\maketitle

\section{Introduction}
\subsection{Background}
\textbf{Transient Response.} The requirement of a good transient response is important in the selection of filters that condition the input signals of a feedback controller like the \gls{pid} \cite{levineSignalProcessingControl2013, hagglundUnifiedDiscussionSignal2013}. This need for filtering is mostly motivated by the need to attenuate measurement or set-point noise fed in as input to a controller. This indirectly connotes the design objectives of smooth tracking (or regulation) and strong filtering
\cite{bavafa-toosiFundamentalLimitations2019}. Often, the controller and the filter are implemented as digital signal processing systems but conveniently designed as analog signal processing systems.

In this context, the oscillatory transient behaviour of most analog filter design forms are not acceptable, as it leads to misinterpretation of the input sequence and therefore inappropriate sequence of control actions. Maximum overshoot is an important and visible index of transient response performance \cite{seronFundamentalLimitationsFiltering1997}. The transient response of signals, in this context, require negligible maximum overshoot. Selection of these filters is therefore, often restricted to the second-order butterworth filter, because of the characteristics of five-percent maximum overshoot, satisfactory noise-reduction, and smaller phase-delay.

Phase-delays represent another problem. However in the case of step-tracking of signals, this delay consideration is sometimes secondary and could be reduced with the extra complexity overhead of lead-lag filters \cite{kennedyRecursiveDigitalFilters2015, hagglundUnifiedDiscussionSignal2013}.

\textbf{All-Pole Transfer Functions.} Interestingly, frequency-selective low-pass filters for noise attenuation are simply all-pole transfer-functions\cite{ribbensSystemsApproachControl2013}. It is well known in the literature that the maximum overshoot is related to the presence of a damping constant in the transfer-function (input--output) response of such filters. However, except for the all-pole second-order standard response, exact values for this damping constant that obtains a defined negligible amount of maximum overshoot from higher-order all-pole transfer-functions, remains unknown.

Linear filtering, a quintessential operation in signal processing and control, can be viewed abstractly as a unity-gain transfer-function mapping \cite{oppenheimDiscreteTimeSignalProcessing2014, pitasNonlinearDigitalFilters1990}. The unity-gain all-pole filter transfer-function design problem can be simplified to specifying only the filter order and a cut-off frequency \cite{carterFastSimpleFilter2018}. Many interesting properties of these transfer-functions are strongly related to polynomial theory. The filtering problem then reduces to using a standard all-pole filter-design form to specify the denominator polynomial of the transfer-function. The denominator polynomial gives the filter's characteristic equation. This then simplifies the filter-design problem to specifying the poles of the filter in terms of the positive real coefficients of the denominator polynomial.

\subsection{Motivation}

\begin{align}
\mathcal{H}\left( \omega \right)&=\begin{cases}
	1&		,\big\lvert \omega \big\rvert <\omega _n\\
	0&		,\text{otherwise}.\\
\end{cases}\label{idealfreqresp}
\end{align}

The behaviour of these standard all-pole filter transfer-function forms is then fully described by the denominator polynomial selected to minimize a certain performance criteria \cite{litovskiElectronicFiltersTheory2019, ribbensSystemsApproachControl2013, seronFundamentalLimitationsFiltering1997}. These standard forms are used to approximate the ideal transfer-function (frequency) response given by (\ref{idealfreqresp}). As the transfer-function order increases, the presence of excessive ringing (oscillations or ripples) becomes visible in the transient-response of these frequency-selective filters. This phenomenon is an important fundamental limitation in many control and filtering applications \cite{bavafa-toosiFundamentalLimitations2019,paarmannDesignAnalysisAnalog2001}.

\begin{align}
\mathcal{H}( \omega )&=\begin{cases}
	1 \quad,0\leqslant \big\lvert \omega \big\rvert  \leqslant ( 1-\zeta ) \omega _n\\
	0 \quad,\big\lvert \omega \big\rvert >( 1+\zeta ) \omega _n\\
    \frac{1}{2} \left[ 1-\sin ( \frac{\pi ( \big\lvert \omega \big\rvert  -\omega _n )}{2\zeta \omega _n} ) \right] \quad,\text{otherwise}.
\end{cases}\label{raisedcosfunc}
\end{align}

\begin{align}
\mathcal{H}\left( \omega \right)&=\begin{cases}
	\frac{1}{2}\left[ 1+\cos \left( \frac{\pi \omega}{2\omega _n} \right) \right]& ,\big\lvert \omega \big\rvert \leqslant 2\omega _n\\
	0& ,\text{otherwise}.
\end{cases}\label{cosbin}
\end{align}

\textbf{Raised cosine functions.} It is known that the study of raised-cosine functions (\ref{raisedcosfunc}) illustrate how much this transient ringings in the time-domain can be damped, while still retaining an approximation to the ideal-filter $(\textstyle{\zeta=0})$ \cite{paarmannDesignAnalysisAnalog2001}.

The raised-cosine function in (\ref{cosbin}) with $\textstyle{\zeta=1}$, is a compact representation of the standard binomial polynomial with no transient overshoot. The raised cosine function (\ref{rcos}) was discussed in \cite{hazonyTimelimitedBandlimitedEnvironment1997}, as an alternative form of binomial expansion. The binomial polynomial is a widely used finite impulse response filter in computer vision and image processing for approximating the gaussian filter function \cite{derpanisOverviewBinomialFilters2005, kupceBinomialFilters1996}. The literature on the use of this polynomial is quite rich. For instance, the binomial window was introduced for interpolating narrow-band signals in \cite{dempsterBinomialWindowHeuristics2000,dempsterLagrangeInterpolatorFilters1999}. \cite{bensouiciSimpleDesignFractional2019} approximated the ideal fractional-delay operator using generalized binomial coefficients.
\begin{align}
\mathcal{H}_n\left( \omega \right)&=\begin{cases}
	\cos ^n \left( \frac{\pi \omega}{2\omega _n} \right)&		,\big\lvert \omega \big\rvert \leqslant 2\omega _n\\
	0&		,\text{otherwise}.
\end{cases}\label{rcos}
\end{align}

Notwithstanding, in \cite{carterFastSimpleFilter2018,porteReplaceDiscreteLowpass2015, ellisFiltersControlSystems2012}, it is claimed that in practise, most filtering require only a unity dc-gain and butterworth response. The \gls{bwf} proposed in \cite{butterworth1930theory}, therefore remains arguably the most widely used among the class of available frequency-selective filters. Although, of slower roll-off (attenuation), the characteristic equation of the butterworth lead to evenly distributed poles in the unit-circle of the normalized complex s-plane, with an added single real pole in the case of odd transfer-function orders \cite{thompsonAnalogLowPassFilters2014, jungStandardResponses2005}.

Further, it was shown in \cite{hazonyTimelimitedBandlimitedEnvironment1997}, that the butterworth filter is a flat-top raised cosine filter. This butterworth transfer-function is regarded as the best achievable transfer-function approximation to the ideal, based on the maximally flat magnitude design criterion in both the pass band and the stop band, among all transfer functions of a given order \cite{bavafa-toosiFundamentalLimitations2019, vandrongelenFiltersDigitalFilters2018, ribbensSystemsApproachControl2013}. They can be represented as cascade of first and second-order polynomials with relative damping constants \cite{dessenOptimizingOrderMinimize2019}. More recently, in \cite{topisirovicUnifiedTheoryStatevariable2015}, the butterworth filter has been classified under a unified theory of \gls{cmac} all-pole filters. The integer-order butterworth filter was generalized to fractional-order in \cite{acharyaExtendingConceptAnalog2014,tsengClosedformDesignsDigital2017}. Also, there are other all-pole transfer-functions detailed in \cite{litovskiElectronicFiltersTheory2019} that outperform the frequency response performance of the butterworth. However, none of them offer a better transient to frequency response compromise like the butterworth. Also, the butterworth filter phase-delay is linear when close to the origin, and is shorter than either the bessel or binomial filter.

Interestingly, the butterworth-filter which shows no ripple in its frequency response bands starts to show considerable ringings in its transient response as the order increases. This flaw corresponds to a poor transient performance index in terms of the maximum overshoot. Consequently, as noted, this constitutes a fundamental limitation in the transient-performance of higher-order transfer-functions designed using the butterworth polynomial. This fundamental fact that the transient performance and sensitivity properties are not consistent with each other, indicates trade-offs inherent in the design of the denominator polynomial. It turns out that the polynomial coefficients of these frequency-selective filters are optimised for frequency response performance at the expense of the transient performance or vice-versa in the case of the bessel filter \cite{paarmannDesignAnalysisAnalog2001} .

\textbf{Damped Binomial Polynomials.} The synthesis of a denominator polynomial with a balanced (good) transient response and frequency response for all positive integer orders is therefore very useful. For higher orders, in connection to the design of the characteristic equation of transfer functions, this problem has been attacked by the use of \gls{bmf} \cite{marchandBinomialSmoothingFilter1983,hernandezAlgorithmsArchitecturesRealTime2000, bennettHistoryControlEngineering1993, tiptonManMachineSystemEqualization1967, oldenburgerOptimalSelfoptimizingControl1966} which directly correspond to (\ref{cosbin}). This denominator polynomial is specified by the use binomial coefficients with uniform damping-constant $\zeta=1$. The binomial filter can be viewed as the upper-limit of the ideal all-pole transfer-function. Therefore, it can be named the Zero-Percent \gls{udbmf}. With its smoother (no overshoot) transient-performance, it poses as a superior design choice for desired transfer functions compared to the butterworth filter and other least-square filters in certain cases \cite{marchandBinomialSmoothingFilter1983}. Although of higher sensitivity, the inherent requirement of no overshoot present (or naturally encoded) in the standard binomial polynomial (real poles only) may not always be a practical choice. It leads to a much slower rise-time in the transient response and a poorer filter-selectivity \cite{bavafa-toosiFundamentalLimitations2019}. A faster transient-response with some form of negligible overshoot is usually preferred. The design of the denominator polynomial in the standard binomial form is therefore limiting.

In applications where both fast, smooth transient-response and strong frequency filtering characteristics are design objectives, we would desire a polynomial representation that is a compromise between the butterworth and binomial standard response, side-stepping the main flaws, while keeping a balanced set of merits, namely: a maximally flat monotonic amplitude in the second-order sense; a quicker roll-off around the cut-off frequency with increasing order; and a faster rise-time with negligible maximum overshoot in the transient response for any order.

\textbf{Non-Uniformly Damped Binomial Polynomials.} Consider the butterworth denominator polynomials (two-decimal place approximated) from (\ref{butfexb})--(\ref{butfexe}). It can be observed that the binomial coefficients (excluding boundary coefficients) are non-uniformly damped for all except the first three cases, where the damping is constant.
{
\fontsize{9}{12}\selectfont
\begin{align}
\mathcal{D}_1(s)&=s+\mathbf{1}\label{butfexb}\\
\mathcal{D}_2(s)&=s^2+2\left( \mathbf{0.71} \right)s +1\label{butfex}\\
\mathcal{D}_3(s)&=s^3 + 3\left( \mathbf{0.67} \right)s^2 + 3\left( \mathbf{0.67 } \right)s + 1\\
\mathcal{D}_4(s)&=s^4+4\left(\mathbf{ 0.65} \right) s^3+6\left( \mathbf{0.57 } \right)s^2 + 4\left( \mathbf{0.65} \right) s+1\\
\mathcal{D}_5(s)&=s^5+5\left( \mathbf{0.65 }\right) s^4+10\left( \mathbf{0.52} \right)s^3 + 10\left( \mathbf{0.52} \right)s^2 + 5\left(\mathbf{0.65}\right)s + 1\\
\mathcal{D}_6(s)&=s^6+6\left( \mathbf{0.64} \right) s^5+15\left( \mathbf{0.49 }\right) s^4+20\left( \mathbf{0.46 }\right) s^3 + 15\left(\mathbf{ 0.49 }\right) s^2+6\left(\mathbf{ 0.64 }\right)s + 1\label{butfexe}%
\end{align}
}
From this, it is clear that the choice of damping constants for coefficients (excluding boundary coefficients) of the binomial polynomial influences the transient-response characteristics of the filter.

\textbf{Main Contributions.} Therefore, in the case of the binomial polynomial as observed from the \gls{bwf} with respect to our filter design objectives, proper choice of damping constants can avoid the flawed transient-response of higher-order cases. Generalizing this design problem, we can impose two possible constraints on the damping constant choice. One, the damping constant should be always uniform in the denominator polynomial, as observed in the standard \gls{bmf} or \gls{bmf0}. Two, for signal reconstruction and control purposes, the desired percentage of maximum-overshoot should be at most five-percent, as realized from (\ref{butfex}). This then reduces the design problem to one question. How exactly should the uniform damping constant $\zeta_n$ be defined for any $n$th-order denominator polynomial?

Therefore, the main goal of this paper is to present a closed-form solution to this question. Given that the uniformly-damped binomial polynomial is optimised on a transient-response criterion. This criterion is the maximum negligible overshoot value observed in the second-order butterworth response.
This paper provides the \emph{explicit closed-form definition of the exact damping-constant that achieves a transient
response with five-percent maximum overshoot for the
uniformly-damped binomial polynomial without the use of explicit
numerical optimization.}

In all, the contributions of this paper is three-fold. \textbf{One}, extension of the binomial expansion theorem to the uniformly-damped binomial theorem. \textbf{Two},definition of the exact damping-constant for a transient response with five-percent maximum overshoot. \textbf{Three}, design and analysis of the \gls{udbmf5}, which more compactly we name \gls{sff} transfer-function.

The rest of this paper is ordered as follows: First, in section~\ref{secudbthm} we start with the introduction of the damped binomial coefficient, and then extend the binomial theorem to the uniformly-damped binomial theorem which leads to the uniformly-damped binomial polynomial. In section~\ref{secudc}, the solution to the exact uniform damping constant that achieves the desired criterion is defined. Table~\ref{table_udbc} illustrates the coefficients (up to the tenth order) of the normalized uniformly-damped binomial polynomial that satisfies the desired transient-response criterion. Definition and analysis of the \gls{udbmf5} transfer-function is presented in section~\ref{secudbtf}. Some numerical performance comparisons of the \gls{udbmf5} to the \gls{bmf} and \gls{bwf} is given in section~\ref{secsims}. Finally, section~\ref{secconcl} concludes the discussion in this paper.

Note that the dynamical system which functions as the frequency-selective filter is represented as a proper transfer function $\mathcal{H}_n$, and its denominator polynomial which is an improper transfer function is represented as $\mathcal{D}$.

\begin{sidewaystable}
\centering
\caption{Five-percent Uniformly-Damped Binomial Filter Transfer Function $\mathcal{H}_n(s)=\frac{1}{\mathcal{D}_n(s)}$}
\label{table_udbc}
\begin{IEEEeqnarraybox}[\IEEEeqnarraystrutmode\IEEEeqnarraystrutsizeadd{2pt}{0pt}]{vx/l/v/c/Vx/c/xv}
\IEEEeqnarraydblrulerowcut\\
&&n&&\zeta_n&&&\IEEEeqnarraymulticol{1}{t}{Polynomial $\mathcal{D}_n(s)$}&&\\
\IEEEeqnarraydblrulerowcut\\
&&1&&1&&&\IEEEeqnarraymulticol{1}{c}{s+\zeta_n}&&\\
\IEEEeqnarraymulticol{10}{h}{}\IEEEeqnarraystrutsize{0pt}{0pt}\\
&&2&&\text{${\sqrt{2}}/{2}$}&&&\IEEEeqnarraymulticol{1}{c}{s^2+2\zeta_n s+1}&&\\
\IEEEeqnarraymulticol{10}{h}{}\IEEEeqnarraystrutsize{0pt}{0pt}\\
&&3&&\text{${\sqrt{5}}/{3}$}&&&\IEEEeqnarraymulticol{1}{c}{s^3+3\zeta_ns^2+3\zeta_ns+1}&&\\
\IEEEeqnarraymulticol{10}{h}{}\IEEEeqnarraystrutsize{0pt}{0pt}\\
&&4&&\text{${\sqrt{10}}/{4}$}&&&\IEEEeqnarraymulticol{1}{c}{s^4+4\zeta_ns^3+6\zeta_ns^2+4\zeta_ns+1}&&\\
\IEEEeqnarraymulticol{10}{h}{}\IEEEeqnarraystrutsize{0pt}{0pt}\\
&&5&&\text{${\sqrt{17}}/{5}$}&&&\IEEEeqnarraymulticol{1}{c}{s^5+5\zeta_ns^4+10\zeta_ns^3+10\zeta_ns^2+5\zeta_ns+1}&&\\
\IEEEeqnarraymulticol{10}{h}{}\IEEEeqnarraystrutsize{0pt}{0pt}\\
&&6&&\text{${\sqrt{26}}/{6}$}&&&\IEEEeqnarraymulticol{1}{c}{s^6+6\zeta_ns^5+15\zeta_ns^4+20\zeta_ns^3+15\zeta_ns^2+6\zeta_ns+1}&&\\
\IEEEeqnarraymulticol{10}{h}{}\IEEEeqnarraystrutsize{0pt}{0pt}\\
&&7&&\text{${\sqrt{37}}/{7}$}&&&\IEEEeqnarraymulticol{1}{c}{s^7+7\zeta_ns^6+21\zeta_ns^5+35\zeta_ns^4+35\zeta_ns^3+21\zeta_ns^2+7\zeta_ns+1}&&\\
\IEEEeqnarraymulticol{10}{h}{}\IEEEeqnarraystrutsize{0pt}{0pt}\\
&&8&&\text{${\sqrt{50}}/{8}$}&&&\IEEEeqnarraymulticol{1}{c}{s^8+8\zeta_ns^7+28\zeta_ns^6+56\zeta_ns^5+70\zeta_ns^4+56\zeta_ns^3+28\zeta_ns^2+8\zeta_ns+1}&&\\
\IEEEeqnarraymulticol{10}{h}{}\IEEEeqnarraystrutsize{0pt}{0pt}\\
&&9&&\text{${\sqrt{65}}/{9}$}&&&\IEEEeqnarraymulticol{1}{c}{s^9+9\zeta_ns^8+36\zeta_ns^7+84\zeta_ns^6+126\zeta_ns^5+126\zeta_ns^4+84\zeta_ns^3+36\zeta_ns^2+9\zeta_ns+1}&&\\
\IEEEeqnarraymulticol{10}{h}{}\IEEEeqnarraystrutsize{0pt}{0pt}\\
&&10&&\text{${\sqrt{82}}/{10}$}&&&\IEEEeqnarraymulticol{1}{c}{s^{10}+10\zeta_ns^9+45\zeta_ns^8+120\zeta_ns^7+210\zeta_ns^6+252\zeta_ns^5+210\zeta_ns^4+120\zeta_ns^3+45\zeta_ns^2+10\zeta_ns+1}&&\\
\IEEEeqnarraydblrulerowcut
\end{IEEEeqnarraybox}
\end{sidewaystable}

\begin{table*}[h]
\centering
\caption{Five-percent Uniformly-Damped Binomial Filter Polynomial Coefficients $\mathcal{H}_n(z)={\mathcal{D}_n(z)}$ (Even $n$ orders only is shown)}
\label{table_udbc}
\fontsize{8}{12}\selectfont
\begin{IEEEeqnarraybox}[\IEEEeqnarraystrutmode\IEEEeqnarraystrutsizeadd{2pt}{0pt}]{vx/l/v/c/Vx/c/xv}
\IEEEeqnarraydblrulerowcut\\
&&n&&\zeta_n&&&\IEEEeqnarraymulticol{1}{t}{Polynomial $\mathcal{D}_n(z)$}&&\\
\IEEEeqnarraydblrulerowcut\\
&&1&&1&&&\IEEEeqnarraymulticol{1}{c}{1\quad 1}&&\\
\IEEEeqnarraymulticol{10}{h}{}\IEEEeqnarraystrutsize{0pt}{0pt}\\
&&2&&\text{${\sqrt{2}}/{2}$}&&&\IEEEeqnarraymulticol{1}{c}{1\quad 2\zeta_n\quad 1}&&\\
\IEEEeqnarraymulticol{10}{h}{}\IEEEeqnarraystrutsize{0pt}{0pt}\\
&&4&&\text{${\sqrt{10}}/{4}$}&&&\IEEEeqnarraymulticol{1}{c}{1\quad 4\zeta_n\quad 6\zeta_n\quad 4\zeta_n\quad 1}&&\\
\IEEEeqnarraymulticol{10}{h}{}\IEEEeqnarraystrutsize{0pt}{0pt}\\
&&6&&\text{${\sqrt{26}}/{6}$}&&&\IEEEeqnarraymulticol{1}{c}{1\quad 6\zeta_n\quad 15\zeta_n\quad 20\zeta_n\quad 15\zeta_n\quad 6\zeta_n\quad 1}&&\\
\IEEEeqnarraymulticol{10}{h}{}\IEEEeqnarraystrutsize{0pt}{0pt}\\
&&8&&\text{${\sqrt{50}}/{8}$}&&&\IEEEeqnarraymulticol{1}{c}{1\quad 8\zeta_n\quad 28\zeta_n\quad 56\zeta_n\quad 70\zeta_n\quad 56\zeta_n\quad 28\zeta_n\quad 8\zeta_n\quad 1}&&\\
\IEEEeqnarraymulticol{10}{h}{}\IEEEeqnarraystrutsize{0pt}{0pt}\\
&&10&&\text{${\sqrt{82}}/{10}$}&&&\IEEEeqnarraymulticol{1}{c}{1\quad 10\zeta_n\quad 45\zeta_n\quad 120\zeta_n\quad 210\zeta_n\quad 252\zeta_n\quad 210\zeta_n\quad 120\zeta_n\quad 45\zeta_n\quad 10\zeta_n\quad 1}&&\\
\IEEEeqnarraydblrulerowcut
\end{IEEEeqnarraybox}
\end{table*}

\section{Uniformly-Damped Binomial Theorem}\label{secudbthm}
In this section, we will introduce the first main result--the uniformly-damped binomial theorem. First, we start by defining the damped binomial coefficients.
\subsection{Damped Binomial Coefficients}\label{secdbcoeff}
\begin{definition}\label{dampcoeff}
Following, the standard definition of binomial (combinatorial) coefficients, for any natural number $n$, the damped binomial coefficient can be written in the form:
\begin{align}
a_i&=\bar{\mathcal{C}}_{i}^{n} \equiv \left( \begin{array}{c}
	n\\
	i\\
\end{array} \right) _{\zeta} \equiv \frac{c}{i!\left( n-i \right) !}
\end{align}
\begin{align}
\noalign{\noindent where, \vspace{\jot}}c &=\begin{cases}
	\zeta \cdot n!& ,0 < i < n\\
	n!&	, i=0 \vee i=n\\
  0& ,n < i < 0\\
\end{cases}
\end{align}
and  $n,i \in \mathbb{N}, n > 0, n \ge i \ge 0$. It follows that, $\bar{\mathcal{C}}_{0}^{n}=\bar{\mathcal{C}}_{n}^{n}=1$, and also $\bar{\mathcal{C}}_{i}^{n}=\bar{\mathcal{C}}_{n-i}^{n}$.
\end{definition}

\subsection{Uniformly-Damped Pascal's Rule}\label{secudpr}
Damped binomial coefficients form the famous universal symmetric pattern, known as the Pascal's triangle. In the uniformly-damped case, this pattern can be defined.
\begin{definition}\label{udpascalaxiom}
The uniformly-damped Pascal's rule is expressed as:
\begin{IEEEeqnarray}{c}
\bar{\mathcal{C}}_{i}^{n+1}=\begin{cases}
	\zeta \cdot \bar{\mathcal{C}}_{i-1}^{n}+\bar{\mathcal{C}}_{i}^{n}& ,i=1\\
	\bar{\mathcal{C}}_{i-1}^{n}+\bar{\mathcal{C}}_{i}^{n}\cdot \zeta& ,i=n\\
	\bar{\mathcal{C}}_{i-1}^{n}+\bar{\mathcal{C}}_{i}^{n}& ,\text{otherwise}.\\
\end{cases}
\end{IEEEeqnarray}
\end{definition}

Next, we will present the uniformly-damped binomial expansion theorem, where the damping constant $\zeta$ is spread uniformly across all of the binomial coefficients, except for the boundary (first and last) coefficients which are not damped.

\subsection{Uniformly-Damped Binomial Polynomial}\label{secudbp}
The denominator polynomial of the all-pole transfer function, can now be defined, by stating the uniformly-damped binomial theorem.
\begin{thm}\label{udbptheorem}
For any natural number $n\in \mathbb{N},n>0$, where $n$ is the order or degree of the polynomial, the uniformly-damped binomial polynomial expansion can be written as:
\begin{IEEEeqnarray}{C}
\mathcal{D}_n=\left( s+\omega _n \right) ^n=\bar{\mathcal{C}}_{0}^{n}s^n+\bar{\mathcal{C}}_{1}^{n}s^{n-1}\omega _n+\bar{\mathcal{C}}_{2}^{n}s^{n-2}\omega _{n}^{2}+\cdots\IEEEnonumber\\
+\bar{\mathcal{C}}_{n-1}^{n}s\omega _{n}^{n-1}+\bar{\mathcal{C}}_{n}^{n}\omega _{n}^{n}=\sum_{i=0}^n{\bar{\mathcal{C}}_{i}^{n}s^{n-i}\omega _{n}^{i}}%
\end{IEEEeqnarray}
\end{thm}
where variables: $s$ is the complex laplace variable, and $\omega_n$ represents the cut-off frequency. The proof of Theorem~\ref{udbptheorem} is shown below.

\begin{proof}
First, consider the normalized form of the binomial expansion
\begin{IEEEeqnarray}{C}
\mathcal{P}\left( n \right) =\left( s+1 \right) ^n=\sum_{i=0}^n{\bar{\mathcal{C}}_{i}^{n}s^{n-i}1^i}=\sum_{i=0}^n{\bar{\mathcal{C}}_{i}^{n}s^{n-i}}\IEEEnonumber\\
\noalign{\noindent From the principle of mathematical induction,  we show that $n = 1$ is true. \vspace
{\jot}}\mathcal{P}\left( 1 \right) =\left( s+1 \right) \equiv \left( s+1 \right)\IEEEnonumber\\
\noalign{\noindent Assume $n=a$ is true. \vspace{\jot}}\mathcal{P}\left( a \right) =\left( s+1 \right) ^a=\sum_{i=0}^a{\bar{\mathcal{C}}_{i}^{a}s^{a-i}}\IEEEnonumber\\
\noalign{ \noindent Then, we show that $n=a+1$ is true. \vspace{\jot}}
\mathcal{P}\left( a+1 \right) =\left( s+1 \right) ^{a+1}=\sum_{i=0}^{a+1}{\bar{\mathcal{C}}_{i}^{a+1} s^{a+1-i}}\IEEEnonumber\\
\noalign{ \noindent Expanding the left-hand side expression, we have: \vspace{\jot}}
\left( s+1 \right) ^{a+1}=\left( s+1 \right) ^a\left( s+1 \right) =\IEEEnonumber\\
\left( \sum_{i=0}^a{\bar{\mathcal{C}}_{i}^{a}s^{a+1-i}} \right) +\left( \sum_{i=0}^a{\bar{\mathcal{C}}_{i}^{a}s^{a-i}} \right)\IEEEnonumber\\
\noalign{\noindent Further expanding and then applying the Damped Pascal rule, \vspace{\jot}} =\bar{\mathcal{C}}_{0}^{a}s^{a+1}+\sum_{i=0}^{a-1}{\left[\bar{\mathcal{C}}_{i+1}^{a}+\bar{\mathcal{C}}_{i}^{a} \right] s^{a-i}}+\bar{\mathcal{C}}_{a}^{a}s^0\IEEEnonumber\\
=\bar{\mathcal{C}}_{0}^{a+1}s^{a+1}+\sum_{i=0}^{a-1}{\bar{\mathcal{C}}_{i+1}^{a+1}s^{a-i}}+\bar{\mathcal{C}}_{a+1}^{a+1}s^0
\IEEEnonumber\\
\noalign{\noindent Shifting the summation index, \vspace{\jot}}= \bar{\mathcal{C}}_{0}^{a+1}s^{a+1}+\sum_{i=1}^a{\bar{\mathcal{C}}_{i}^{a+1}s^{a+1-i}}+\bar{\mathcal{C}}_{a+1}^{a+1}s^0=
\sum_{i=0}^{a+1}{\bar{\mathcal{C}}_{i}^{a+1}s^{a+1-i}}\IEEEnonumber%
\end{IEEEeqnarray}
Thus, we have shown by inductive hypothesis that since the damped binomial theorem is true for any arbitrary natural number $a$, then it is true for $a+1$
that is $\mathcal{P}\left( a \right) \rightarrow  \mathcal{P}\left( a+1 \right)$.
As $\mathcal{P}\left( 1 \right)$ is true, it follows therefore from mathematical induction that $\mathcal{P}\left( n \right)$ is true for all natural numbers and so the theorem is established.  Since the normalized expression is true, the denormalised expression also follows as true. This is proved by noting that
$$
\left( s+\omega _n \right) ^n=\omega _{n}^{n}\left( \frac{s}{\omega _n}+1 \right) ^n
$$
\begin{IEEEeqnarray}{rCl}
&=\omega ^n\left[\bar{\mathcal{C}}_{0}^{n}\left( \frac{s}{\omega _n} \right) ^n+\sum_{i=1}^{n-1}{\bar{\mathcal{C}}_{i}^{n}\left( \frac{s}{\omega _n} \right) ^{n-i}} + \bar{\mathcal{C}}_{n}^{n}\left( \frac{s}{\omega _n} \right) ^0 \right]\IEEEnonumber%
\end{IEEEeqnarray}
$$
=\bar{\mathcal{C}}_{0}^{n}s^n+\sum_{i=1}^{n-1}{\bar{\mathcal{C}}_{i}^{n}s^{n-i}\omega _{n}^{i}}+\bar{\mathcal{C}}_{n}^{n}\omega _{n}^{n}=\sum_{i=0}^n{\bar{\mathcal{C}}_{i}^{n}s^{n-i}\omega _{n}^{i}}
$$
\end{proof}

\begin{axiom}\label{udbpsum}
The sum of the coefficients in a uniformly-damped binomial polynomial is: $2 + \left( 2^{n}-2 \right)\zeta$, $\forall n>0$.
\end{axiom}

Recall, we desire to achieve a maximum overshoot less or equal to that of the second-order butterworth filter (\gls{bwf}). At this point, the problem is to determine, for each order, what the damping constant applied uniformly to the binomial coefficients will exactly be in numeric value. This could be solved by employing numerical optimization, however in this paper, we define a simpler, direct and straight-forward closed-form expression that solves this problem.

\subsection{Uniform-Damping Constant}\label{secudc}
The core result of this paper lies in the explicit definition of the uniform-damping constant solution to the five-percent maximum overshoot filtering design using a binomial polynomial.
\begin{figure}[!t]
  \centering
  \includegraphics[scale=0.55]{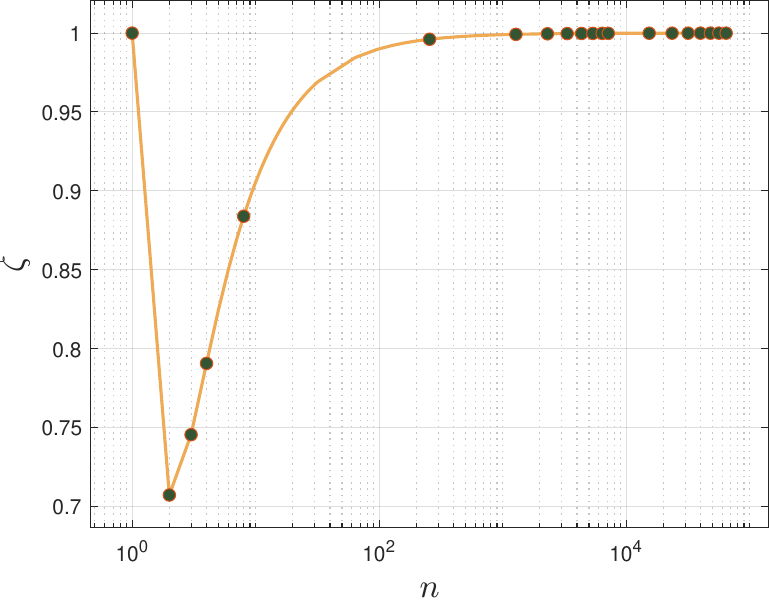}
  \caption{Semi-log plot showing the behaviour, as $n\to\infty$, of the uniform damping-constant $\zeta$ that satisfies the $M_p\le 5\%$ constraint}\label{dampfeats}
\end{figure}

Consider the $n$th order uniformly-damped binomial polynomial in Theorem~\ref{udbptheorem}. The exact uniform-damping constant that satisfies the maximum overshoot optimization criterion $M_p$ of the second-order \gls{bwf} can be defined as follows:
\begin{axiom}\label{exactdmpaxiom}
The uniform-damping constant that satisfies a $M_p\le 5\%$, for the $n$th order uniformly-damped binomial polynomial in Theorem~\ref{udbptheorem} is:
\begin{IEEEeqnarray}{c}
\zeta =\zeta _n=\frac{\sqrt{n\left( n-1 \right) -\left( n-2 \right)}}{n}\label{exactdamping}%
\end{IEEEeqnarray}
\end{axiom}

The result or expression in (\ref{exactdamping}) represents the solution to the ringing problem present in the specification of damped binomial polynomial representations such as the \gls{bwf}. It constrains the $M_p$ to be no more than five-percent. We note that the design choice of five-percent $M_p$ in this paper is because it is the most common allowable $M_p$ in filtering for control tasks \cite{fadaliDigitalControlEngineering2020,boltonSystemResponse2021}.

Also, it can be observed that for large $n$, the numerator of (\ref{exactdamping}) is $\sqrt{(n^{2} - {2\,n} + 2)} \approx n$. This implies that for very large $n$ in the limit such as $n>=2^{8}$, the damping value becomes numerically equivalent to $1$. More, formally, as $n\to\infty$, then $\zeta\to1$. That is as illustrated in (Fig.~\ref{dampfeats}), for $n>1$, as n is increased, the value of $\zeta$ gradually moves from $\frac{\sqrt{2}}{2}$ back to 1.

\begin{figure}[!t]
\centering
\subfloat[]{\includegraphics[scale=0.8]{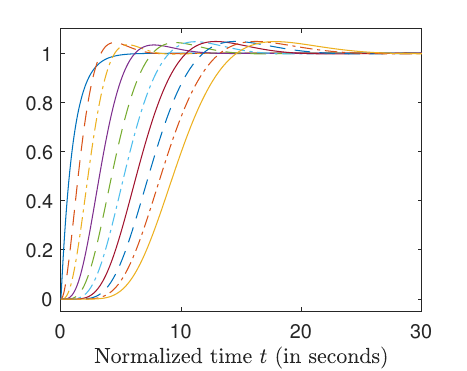}%
\label{figstep}}
\quad
\subfloat[]{\includegraphics[scale=0.8]{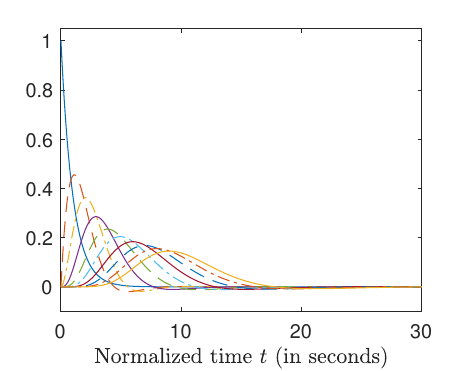}%
\label{figimp}}
\caption{Step (\ref{figstep}) and Impulse Response (\ref{figimp}) plot of the uniformly-damped binomial low-pass filter transfer-function with normalized cut-off frequency for values of $n=1$ (blue) to $n=10$ (brown). }
\label{figtransresp}
\end{figure}

Axiom~\ref{exactdmpaxiom} can be proved using the $M_p$ formula: $\exp\left(\frac{\zeta\pi}{\sqrt{1-\zeta^2}}\right)\times 100\%$ for a second-order prototype system response. Also, this is empirically illustrated of the transient step and impulse response of the \gls{udbmf5} transfer-function in (Fig.~\ref{figtransresp}).
The synthesis of the \gls{udbmf5} is discussed in the next section.

\section{Uniformly-Damped Binomial Filter (Transfer-Function)}\label{secudbtf}

Applying Theorem~\ref{udbptheorem} and Axiom~\ref{exactdmpaxiom} to the synthesis of the unity-gain continuous-time transfer-function, where $k_0=1$. The uniformly-damped binomial filter $\mathcal{H}_n$ is defined as:
\begin{IEEEeqnarray}{Rl}
\mathcal{H}_n\left( s \right)=&\frac{k_0\,\omega _{n}^{n}}{\mathcal{D}_n(s)}=\frac{k_0}{\sum_{i=0}^n{\bar{\mathcal{C}}_{i}^{n}\left( s/\omega _n \right) ^{n-i}}}\label{udbf}%
\end{IEEEeqnarray}

\begin{figure}[thpb]
\centering
\subfloat[]{\includegraphics[scale=0.8]{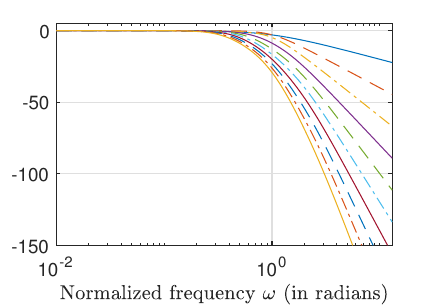}%
\label{figmagdb}}
\quad
\subfloat[]{\includegraphics[scale=0.8]{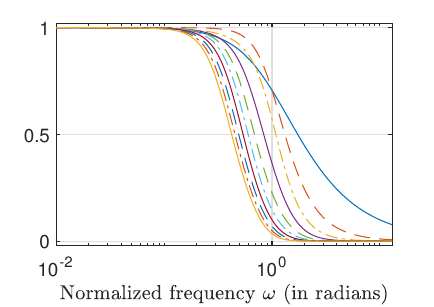}%
\label{figmagn}}
\caption{Magnitude Response Plots of the \gls{udbmf5} with normalized cut-off frequency for orders $n=1$ (blue) to $n=10$ (brown): shown in decibels in (\ref{figmagdb}) compared to (\ref{figmagn}). }
\label{figmagresp}
\end{figure}

The expressions in (\ref{magf}) and (\ref{magsqf}) respectively give the magnitude, and squared-magnitude of the \gls{udbmf5}.
\begin{IEEEeqnarray}{C}
\big\lvert \mathcal{H}_n\left( \omega \right) \big\rvert =\frac{1}{\sqrt{\left( {\omega}/{\omega _n} \right) ^{2n}+\kappa +1}}\label{magf}\\
\big\lvert \mathcal{H}_n\left( \omega \right) \big\rvert ^2=\mathcal{H}_n\left( s \right) \mathcal{H}_n\left( -s \right) =\frac{1}{\left( {\omega}/{\omega _n} \right) ^{2n}+\kappa +1}\label{magsqf}\\
\kappa =\sum_{i=n-1}^1{\alpha _t\left( {\omega}/{\omega _n} \right) ^{2i}}\\
\alpha _t=\left( \bar{\mathcal{C}}_{t}^{n} \right) ^2+2\sum_{r=1}^{\bar{r}}{\left( -1 \right) ^r\bar{\mathcal{C}}_{j}^{n}\bar{\mathcal{C}}_{k}^{n}}\\
\noalign{\noindent and \vspace{\jot}}\bar{r}=\begin{cases}
	n-i&		,\begin{cases}
	i\geqslant \frac{n}{2}\,\,	\text{and even $n$}\\
	i\geqslant \frac{\left( n-1 \right)}{2}\,\,\text{and odd $n$}\\
\end{cases}\\
	i&		,\text{otherwise}.\\
\end{cases}\IEEEnonumber%
\end{IEEEeqnarray}
where, $t=n-i$, $j=t-r$, $k=t+r$ and $\alpha _t=\alpha _{n-t}$.
\begin{figure}[!tb]
\centering
\subfloat[]{\includegraphics[scale=0.7]{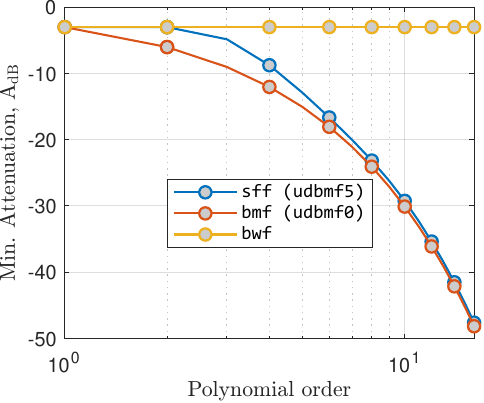}%
\label{figattdb}}
\quad
\subfloat[]{\includegraphics[scale=0.7]{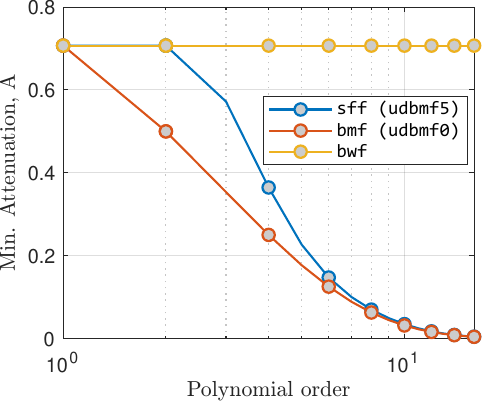}%
\label{figatt}}
\caption{The Minimum Attenuation at the normalized cut-off frequency for orders $n=1$ to $n=16$: shown in decibels in (a.) and in dimensionless units in (b.) }
\label{figattplot}
\end{figure}
Also, for any order $n$, the minimum attenuation at a given frequency is given by (\ref{minattf}), and the bandwidth given a minimum attenuation can be obtained by solving the polynomial equation (\ref{bwf}). From inspection of the slope of the magnitude (in dB) plot in Fig.~\ref{figmagdb}, as $\omega >> \omega_n$, it is clear that high-frequency roll-off is $-20\,n$ dB/decade.
\begin{IEEEeqnarray}{c}
A_{\mathrm{dB}}=\text{10}\log \left( \left( {\omega}/{\omega _n} \right) ^{2n}+\kappa+1 \right)\label{minattf}\\
\left( \omega/\omega _n \right) ^{2n} + \kappa +\left( 1-10^{A_{\text{dB}}/10} \right)=0\label{bwf}%
\end{IEEEeqnarray}

The minimum attenuation at the normalized cut-off frequency for increasing orders of the binomial filters under consideration in this work are illustratively compared in Fig.~\ref{figattplot}. It can be observed that of the three while the \gls{bwf} has a constant and maximum $A_\mathrm{dB}$ (-3dB or 0.7071), the \gls{udbmf5} gives the median $A_\mathrm{dB}$ and approaches that of the \gls{bmf0} having the smallest $A_\mathrm{dB}$ at the cut-off frequency, as $n$ increases in value.

The \gls{udbmf5}  becomes a digital infinite impulse response (IIR) filter by $s\to z$ bilinear transformation. Also, the denominator polynomial can be directly applied as a digital finite impulse response (FIR) filter (sliding window filter) by directly replacing $s=z$ and $\omega_n=1$ as shown in (\ref{zfir}), having $N=n+1$ coefficients or components.
\begin{IEEEeqnarray}{c}
\mathcal{D}\left( z \right) =\left( z+1 \right) ^n=\sum_{i=0}^n{\bar{\mathcal{C}}_{i}^{n}z^{n-i}}\label{zfir}
\end{IEEEeqnarray}
\subsection{Flatness and Selectivity of Filter}\label{secflats}
\begin{figure}[!t]
\centering
\subfloat[]{\includegraphics[scale=0.65]{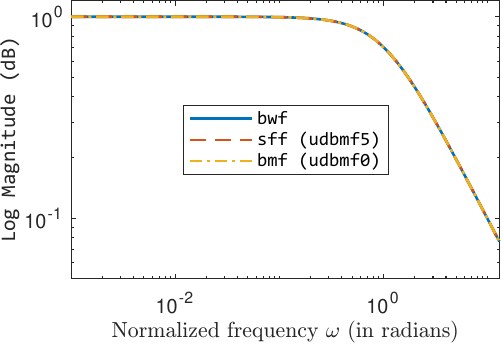}%
\label{fign1}}
\quad
\subfloat[]{\includegraphics[scale=0.65]{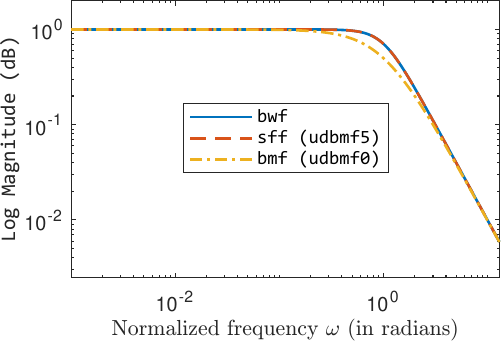}%
\label{fign2}}
\quad
\subfloat[]{\includegraphics[scale=0.65]{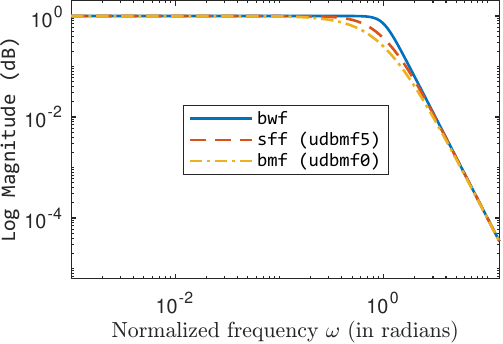}%
\label{fign3}}
\quad
\subfloat[]{\includegraphics[scale=0.65]{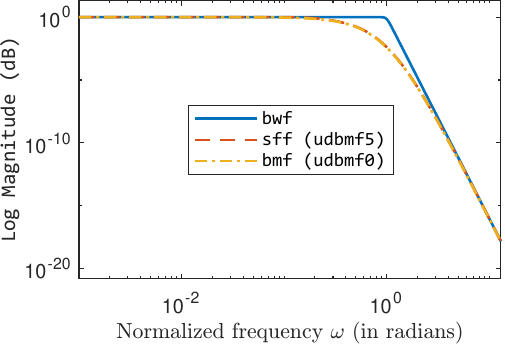}%
\label{fign4}}
\caption{Magnitude Response comparison with normalized cut-off frequency for orders (a.) $n=1$ (b.) $n=2$ (c.) $n=4$ and (d.) $n=16$ }
\label{figcmpmag}
\end{figure}

Given the magnitude of the filter, the magnitude flatness and selectivity of the filter can be investigated by respectively finding the derivative of the magnitude with respect to the frequency and the negative derivative of the magnitude with respect to the frequency at the origin.
\begin{IEEEeqnarray}{C}
\frac{\text{d}\big\lvert \mathcal{H}_n\left( \omega \right) \big\rvert }{\text{d}\omega}=-\big\lvert \mathcal{H}_n\left( \omega \right) \big\rvert ^3\left( \frac{n}{\omega _n}\left( \frac{\omega}{\omega _n} \right) ^{2n-1}+\frac{\text{d}\kappa}{\text{d}\omega} \right)\label{dfdw}\\
\text{where, }\frac{\text{d}\kappa}{\text{d}\omega}=\sum_{i=n-1}^1{\alpha _t\frac{i}{\omega _n}\left( \frac{\omega}{\omega _n} \right) ^{2i-1}}
\end{IEEEeqnarray}
From (\ref{dfdw}), it is easy to see that, since (\ref{magf}) is positive, the derivative monotonically decreases with no ripple. The $k$th, ($k = 1, ... \infty$) derivatives of the gain are zero
at the origin ($\omega=0$) of the closed left-half complex plane for all $n$, except for even indexed $k$th derivatives, $k= 2i$ up to $2n$, where $i= 1,2,...,n$, provided $n > 2$. As illustrated in Fig.~\ref{figcmpmag}, this results in a maximal flatness for $n <=2$, a flatness in between that of the \gls{bwf} and \gls{bmf0} for $n > 2$, and a flatness approximately that of the \gls{bmf0} as $n$ increases in the allowable finite integer limit.
\begin{equation}
-\frac{\text{d} \lvert \mathcal{H}_n\left(\omega\right) \rvert }{\text{d}\omega} \bigg\rvert_{\omega =\omega _n}=\frac{\frac{n}{\omega _n}+\sum_{i=n-1}^1{\alpha _t\frac{i}{\omega _n}}}{\left[ 2+\sum_{i=n-1}^1{\alpha _t} \right] ^{\frac{3}{2}}}\label{fs}%
\end{equation}
It is obvious from (\ref{fs}), that the frequency selectivity of \gls{udbmf5} is the same as that of the \gls{bwf} for $n<=2$, it is between that of the \gls{bwf} and \gls{bmf0} for $n>2$, and then it is approximately the same as that of the \gls{bmf0} for $n\to\inf$.

\subsection{Phase-Delay and Group-Delay}\label{secpgdelay}
\begin{figure}[thpb]
\centering
\includegraphics[scale=0.8]{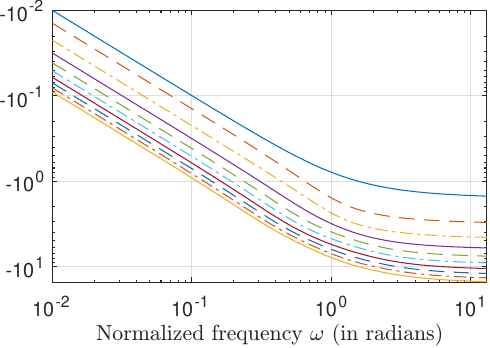}%
\caption{Phase Response Plot (in radians) of the uniformly-damped binomial low-pass filter with normalized cut-off frequency for orders $n=1$ (blue) to $n=10$ (brown). }
\label{figphaseresp}
\end{figure}

The phase response of the \gls{udbmf5} is illustrated in (Fig.~\ref{figphaseresp}). The linear response can be investigated through the expressions (\ref{pdelay}) for the phase-delay and (\ref{gdelay}) for the group-delay.

\begin{IEEEeqnarray}{R}
\tau _p\left( \omega \right) =\frac{\omega _n}{\omega}\text{arctan}\left( \frac{\sum_{i=\text{odd}}^n{\left( -1 \right) ^{\frac{\left( i-1 \right)}{2}}\bar{\mathcal{C}}_{i}^{n}\left( \frac{\omega}{\omega _n} \right) ^i}}{\sum_{i=\text{even}}^n{\left( -1 \right) ^{\frac{i}{2}}\bar{\mathcal{C}}_{i}^{n}\left( \frac{\omega}{\omega _n} \right) ^i}} \right)\label{pdelay}
\end{IEEEeqnarray}
\begin{IEEEeqnarray}{c}
\tau _g\left( \omega \right) =\big\lvert \mathcal{H}_n\left( \omega \right) \big\rvert ^2\left( \left( {\omega}/{\omega _n} \right) ^{2n-2}+\delta +n \right)\label{gdelay}\\
\delta =\sum_{i=n-2}^1{\lambda _t}\left( {\omega}/{\omega _n} \right) ^{2i}\\
\lambda _t=\sum_{r=1}^{\bar{r}}{\begin{array}{c}
	\left( -1 \right) ^{r-1}\bar{\mathcal{C}}_{j}^{n}\bar{\mathcal{C}}_{k}^{n}\\
\end{array}}\\
\noalign{\noindent and, \vspace{\jot}}\bar{r}=\begin{cases}
	n-i&		,\begin{cases}
	i\geqslant \frac{n}{2}\,\, \text{and even $n$}\\
	i\geqslant \frac{\left( n-1 \right)}{2}\,\, \text{and odd $n$}\\
\end{cases}\\
	i+1&		,\text{otherwise}.\\
\end{cases}\IEEEnonumber%
\end{IEEEeqnarray}
where $t=n-1-i$, $j=t+1-r$, $k=t+r$ and $\lambda _t=\lambda _{n-t-1}$.

\begin{figure}[thpb]
\centering
\includegraphics[scale=0.8]{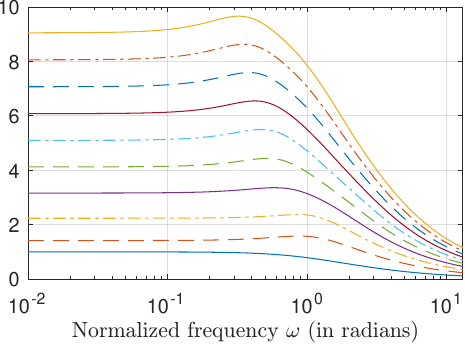}%
\caption{Phase Delay of the uniformly-damped binomial low-pass filter with normalized cut-off frequency for orders $n=1$ (blue) to $n=10$ (brown). }
\label{figtaup}
\end{figure}

\begin{figure}[thpb]
\centering
\includegraphics[scale=0.8]{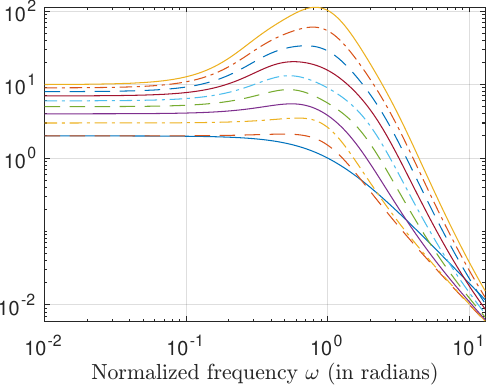}%
\caption{Group Delay (in seconds) of the uniformly-damped binomial low-pass filter with normalized cut-off frequency for orders $n=1$ (blue) to $n=10$ (brown). }
\label{figtaug}
\end{figure}

At $\omega=\omega_n$, the total phase in radians is $\frac{n\pi}{4}$. At the origin $\omega=0$, the phase-delay (Fig.~\ref{figtaup}) seen at the output of the filter is $\frac{\zeta\,n}{\omega_n}$ and the group-delay (Fig.~\ref{figtaug}) seen at the output of the filter is $2$ for $n=1$ and $\frac{n}{\omega_n}$ for $n>1$. Both delays are exactly proportional to the order of the filter, and increase nonlinearly as the order is increased. Notably, we see that with the help of the damping constant the phase-delay of the \gls{udbmf5} is less than the standard \gls{bmf} (\gls{bmf0}).

The transient and frequency response analysis of the \gls{udbmf5} poses it as a balanced compromise between the butterworth and binomial standard forms. It provides a balance of: excellent transient performance in its time-response (Fig.~\ref{figtransresp}) with a good sensitivity and selectivity performance across the passband and stopband in its frequency response (Fig.~\ref{figmagresp} and Fig.~\ref{figphaseresp}).

\section{Simulation}\label{secsims}
In this section, we compare the \gls{udbmf5}, with other related binomial filters: \gls{bwf}, \gls{bmf0}. In the FIR case, we also compare the binomial filters with a cubic least-square polynomial fitted \gls{sgf}. The implemented filter specs are of even filter polynomial order $n$, implying symmetric odd-number of coefficients $N$, and hence ideal for smoothing applications \cite{marchandBinomialSmoothingFilter1983}.
\subsection{Case 1: Online Signal Filtering}
\begin{figure}[thpb]
\centering
\subfloat[]{\includegraphics[scale=0.6]{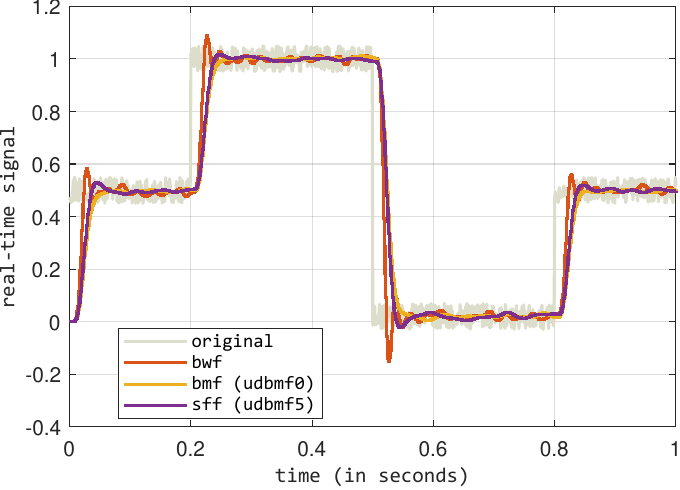}%
\label{figcmpiirone}}
\quad
\subfloat[]{\includegraphics[scale=0.6]{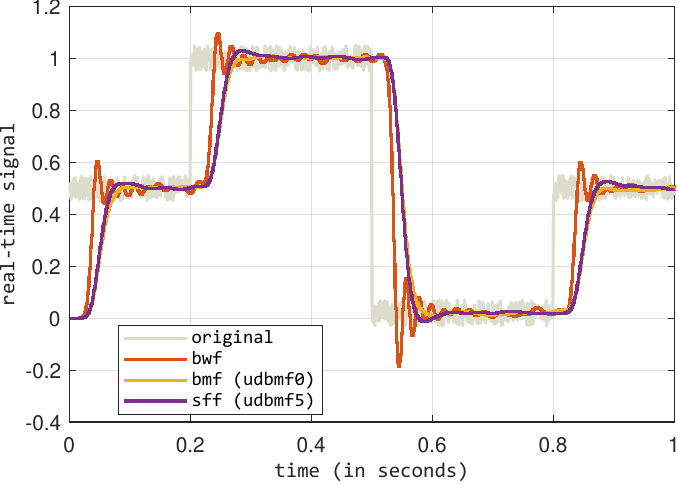}%
\label{figcmpiirtwo}}
\caption{IIR filtered output response of a real-time tracked noisy step signal. \textbf{(a.)}: 8th-order. \textbf{(b.)}: 16th-order.}
\label{figonone}
\end{figure}

In application to higher-order digital infinite impulse response filtering of a noisy signal, with negligible overshoot, the \gls{udbmf5} is contrasted to both the \gls{bwf} and standard \gls{bmf0} as illustrated in (Fig.~\ref{figcmpiirone} and Fig.~\ref{figcmpiirtwo}).
\subsection{Remarks}
It can be distinctly observed from the signal waveforms in Fig.~\ref{figonone} that the \gls{udbmf5} (or \gls{sff}) is a compromise response with respect to the other two binomial filter types, useful where fast, smooth transient response and strong filtering are both required for the binomial filter polynomial orders. This is the core motivation and context for designing the \gls{udbmf5}.
\subsection{Case 2: Offline Data Filtering}
\subsubsection{Case 2.1: Noisy Pulse Data}
In many cases, logged raw sensor data, for further system design and analysis are used to perform more experiments or to make predictions. In this case, we consider a logged synthetic data containing normalized speed values of a dc-motor obtained via a sensor sampled at $1$ms. For, clarity, we only show a section of the offline pulse sequence between the time-window of the $60$th and $120$th index.

\begin{figure}[thpb]
\centering
\subfloat[]{\includegraphics[scale=0.5]{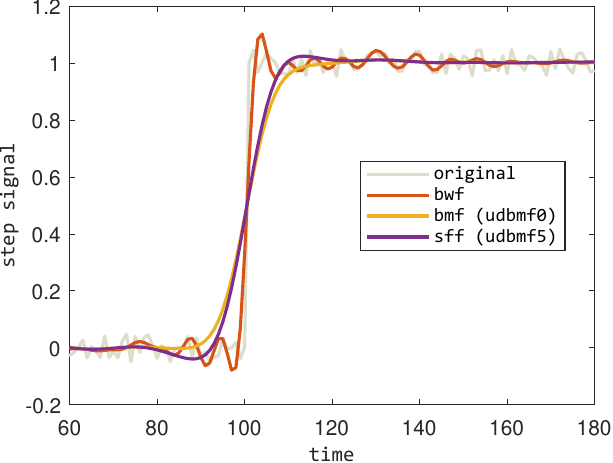}%
\label{figopulsea}}
\quad
\subfloat[]{\includegraphics[scale=0.5]{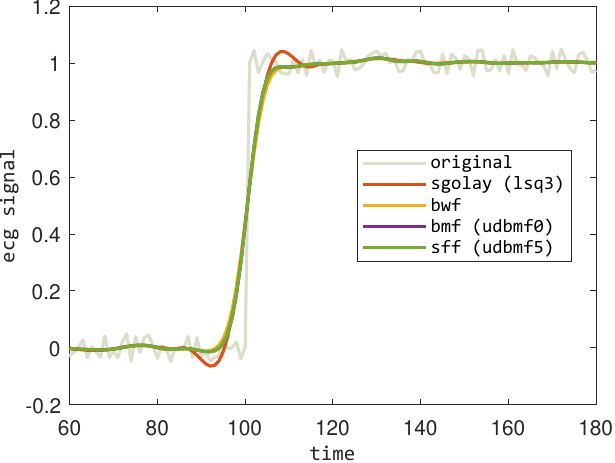}%
\label{figopulseb}}
\vspace{1ex}
\subfloat[]{\includegraphics[scale=0.5]{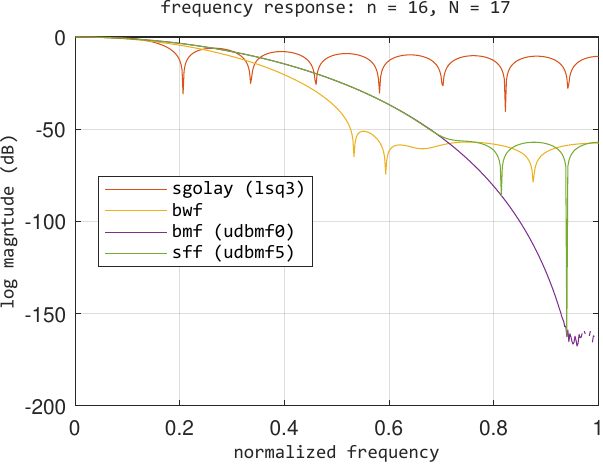}%
\label{figopulsec}}%
\quad
\subfloat[]{\includegraphics[scale=0.8]{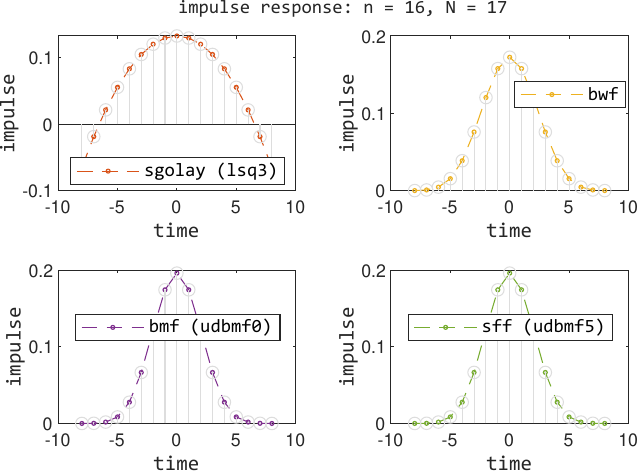}%
\label{figopulsed}}%
\caption{$16$th-order filtering of a noisy pulse data. \textbf{(a.)} IIR output, \textbf{(b.)} FIR output, \textbf{(c.)} Frequency response, \textbf{(d.)} Impulse response.}
\label{figoffsense}
\end{figure}

We illustrate and compare the behaviour and response of four filter classes for a polynomial order of $16$: low pass filtering in Fig.~\ref{figoffsense}.
\subsubsection{Case 2.2: Noisy ECG Data}
The electrocardiography (ECG) signals (electrical activity of the heart) play a key role in diagnosing diverse kinds of cardiovascular activities and diseases. Extracting certain features from raw ECG data is common for bio-informatics and bio-medical purposes. Logged or recorded ECG data are usually noisy, as such it is known that frequency selective filters must balance both smoothing and signal preservation, without distorting accurate diagnosis and interpretation.
In a very noisy case, such is useful in detecting, for example the \textbf{QRS wave complex} feature in ECG signals, which is further used to make diagnostic predictions of the healthy state of the heart \cite{lastre-dominguezECGSignalDenoising2019}. Typically for healthy adults, this is a short-duration electrical impulse activity lasting between $0.06$s and $0.10$s.

In this case, we consider a synthetic Electrocardiogram (ECG) signal data waveform of length $500$ and sampled at a frequency of $1$kHz from an ECG data acquisition and monitoring device. The following illustrates zero-phase filtering of the synthetic electrocardiogram (ECG) waveform in order to smooth and also preserve features exactly where they occur in the unfiltered signal. In this case, the electrical impulse begins, as seen from the time-window at around time $0.16$s and ends at about $0.22$s, which is an average duration of $0.06$s.

\begin{figure}[thpb]
\centering
\subfloat[]{\includegraphics[scale=0.5]{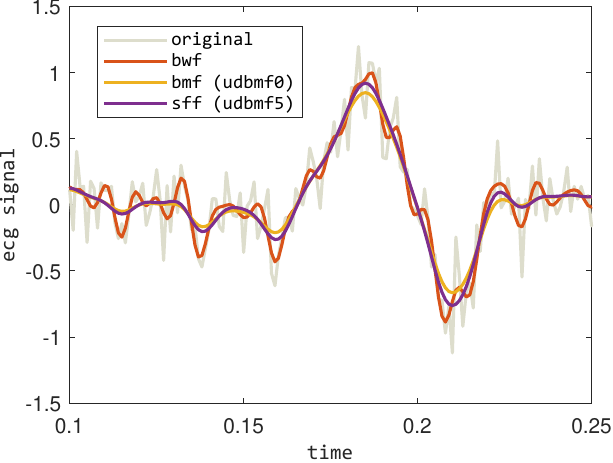}%
\label{figecgba}}
\quad
\subfloat[]{\includegraphics[scale=0.5]{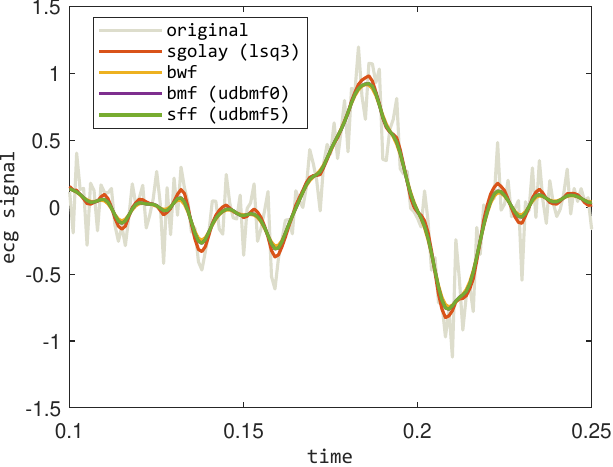}%
\label{figecgbb}}
\vspace{1ex}
\subfloat[]{\includegraphics[scale=0.6]{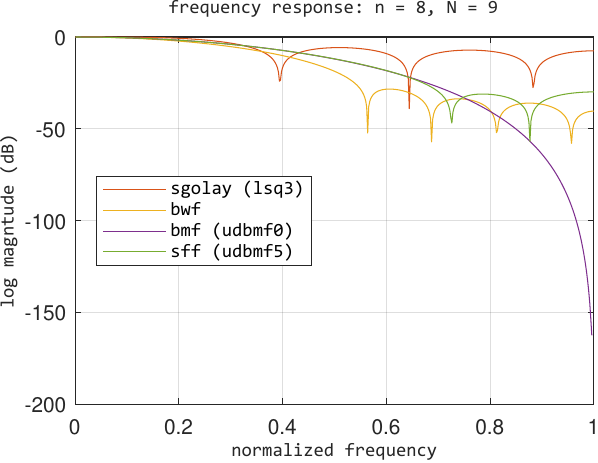}%
\label{figecgbc}}%
\quad
\subfloat[]{\includegraphics[scale=0.8]{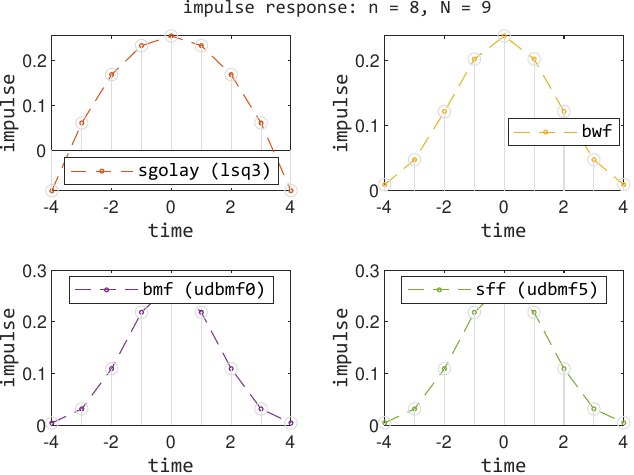}%
\label{figecgbd}}%
\caption{$8$th-order filtering of a ECG data. \textbf{(a.)} IIR output, \textbf{(b.)} FIR output, \textbf{(c.)} Frequency response, \textbf{(d.)} Impulse response.}
\label{figoffecgb}
\end{figure}

We illustrate and compare the behaviour and response of four filter classes for a polynomial order of $8$: low pass filtering in Fig.~\ref{figoffecgb}.

\begin{figure}[thpb]
\centering
\subfloat[]{\includegraphics[scale=0.5]{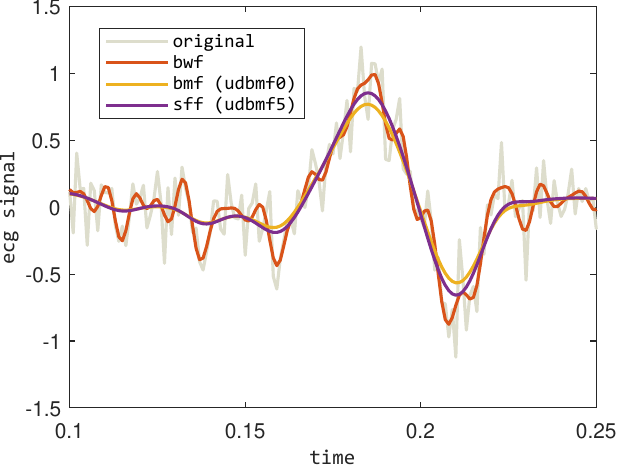}%
\label{figecgaa}}
\quad
\subfloat[]{\includegraphics[scale=0.5]{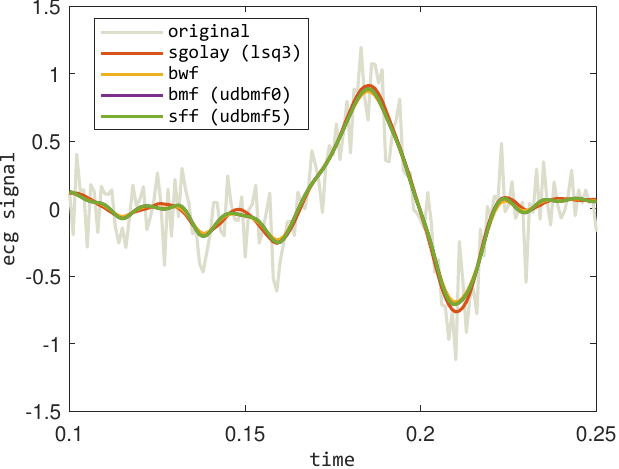}%
\label{figecgab}}
%
\vspace{1ex}
\subfloat[]{\includegraphics[scale=0.6]{offstepfreqr16.pdf}%
\label{figecgac}}%
\quad
\subfloat[]{\includegraphics[scale=0.8]{offstepimpr16.pdf}%
\label{figecgad}}%
\caption{$16$th-order filtering of a ECG data. \textbf{(a.)} IIR output, \textbf{(b.)} FIR output, \textbf{(c.)} Frequency response, \textbf{(d.)} Impulse response.}
\label{figoffecga}
\end{figure}

Further, we again illustrate and compare the behaviour and response of four filter classes for a polynomial order of 16: low pass filtering in Fig.~\ref{figoffecga}.

\subsection{Remarks}
Generally, it can be observed from the presented simulated case-examples that although with good attenuation characteristics, in the IIR case, the \gls{bwf} distorts the real-time signal badly. Meanwhile the overshoot is just once and minimal to within five-percent for the \gls{sff}, while the \gls{bmf0} shows no overshoot.
For the FIR case, we consider the \gls{sgf} filter, and we see that like the \gls{bwf}, the use of long sequences of \gls{sgf} polynomials lead to problems like overshoots, which in our application of interest is not desirable.
In comparison, we observe that the frequency-response of the binomial polynomials give lesser overshoots and more noise attenuation. The impulse response clearly shows that \gls{bwf}, \gls{bmf0} and \gls{sff} are binomial filter polynomials.
\section{Conclusions}\label{secconcl}
In this paper, we have introduced and preliminarily discussed a new class of binomial filter (or transfer-function) design. It was first shown that the butterworth filter is a uniformly damped binomial filter for lower orders, and then non-uniformly damped binomial filter for higher orders. In contrast this paper by extending the binomial theorem and providing an explicit uniform damping constant formula introduced the five-percent uniformly-damped binomial filter \gls{sff} or \gls{udbmf5}. The basis is a uniform-damping constant optimized on the popular $5\%$ maximum-overshoot criterion of the second-order butterworth filter. This class of filter or standard form represent a compromise of the strong merits of both the butterworth and the standard binomial filter. The filter meets all imposed constraints (time-invariant, causal, linear, proper rational transfer function of finite order, with positive real coefficients given by the uniformly-damped binomial polynomial) that assure the realization of a practical analog or digital filter. Clearly, in terms of simplicity and computational effort, the \gls{bmf0} (integer-valued separable polynomial coefficients) supersedes the \gls{udbmf5} (real-valued polynomial coefficients). In addition, for higher orders, like the \gls{bwf}, IIR realization of the \gls{udbmf5} is affected by rounding-error accumulation due to finite-precision machines.

Potential applications of this filter include: the design of linear output-state estimators with their derivatives, and the design of characteristic equations for higher-order control systems with balanced transient response (no ringing) and frequency filtering objectives.

One important limitation, of the \gls{udbmf5} in this work, is that unlike the butterworth filter \gls{bwf} and standard binomial filter \gls{bmf0}, there is currently no exact closed-form formula for determining the pole positions of the five-percent uniformly-damped binomial filter. Also, except for order $n<=2$, the filter under consideration has no unique roots or separable convolution kernels, that is higher order sequences of the filter polynomial cannot be elegantly decomposed into a combination of lower order forms. This problem poses as a interesting and challenging future work, which readers may attempt to disproof. Possibly, in a future work, we will more explicitly describe this class of binomial filters and its applications in offline signal processing, for example as n-dimensional smoothing kernels, differentiating filter kernels, and so on. Please see the Github repository \cite{somefunSFF2021} for the algorithms and scripts used for producing the results in this paper.

\backmatter


\section*{Declarations}

\bmhead{Conflict of interest}

The authors declare that they have no conflict of interest.

%

\bmhead{Funding}

This work was not funded by any organization

\bmhead{Code availability}

The code used to develop this article can be found on Github \url{https://github.com/somefunagba/sff}.

%

%
%
%
%

\bibliography{DBF}

\end{document}